\documentclass[aps,prl,eqsecnum,showpacs,preprint]{revtex4}

\usepackage{ulem}
\usepackage{amsfonts}
\usepackage{amsmath}
\usepackage{amssymb}
\usepackage{graphicx}
\begin{document}

\title{Electrical Spin Injection in Multi-Wall carbon NanoTubes with transparent
ferromagnetic contacts}

\author{S. Sahoo, T. Kontos and C. Sch\"{o}nenberger}
\affiliation{Institut f\"{u}r Physik, Universit\"{a}t Basel, Klingelbergstrasse 82,\\
CH-4056 Basel, Switzerland.}
\author{C. S\"urgers}
\affiliation {Physicalishes Institut and DFG Center for Functional
Nanostructures, University of Karlsruhe, D-76128 Karlsruhe,
Germany}

\date{\today}


\begin{abstract}
We report on electrical spin injection measurements on Multi-Wall
carbon NanoTubes (MWNTs) . We use a ferromagnetic alloy
Pd$_{1-x}$Ni$_{x}$ with x $\approx$ 0.7 which allows to obtain
devices with resistances as low as $5.6$ $k\Omega$ at $300$ $K$.
The yield of device resistances below $100$ $k\Omega$, at $300$
$K$, is around 50\%. We measure at $2$ $K$ a hysteretic
magneto-resistance due to the magnetization reversal of the
ferromagnetic leads. The relative difference between the
resistance in the antiparallel (AP) orientation and the parallel
(P) orientation is about $2\%$.
\end{abstract}

\maketitle

How the spin degree of freedom propagates and can be manipulated
in low dimensional devices is a question of both fundamental and
technical interest. On one hand, proposals for a spin field-effect
transistor (SpinFET) \cite{Datta:90}, which one can consider as a
generic spintronic scheme, rely on electrical spin injection in
1-dimensional channels. On the other hand, spin transport is
expected to provide new information on the peculiar nature of an
electronic fluid, as electron-electron interactions are enhanced
when one reduces the dimensionality. Within the framework of the
Luttinger liquid model, for example, Balents and Egger showed
theoretically that spin-charge separation modifies qualitatively
spin transport in quantum wires \cite{Egger:01}.

Carbon nanotubes (NTs) can be considered as 1-dimensional or
0-dimensional conductors, with important Coulomb interaction
\cite{Dekker:98,Smalley:99}. Spin transport is thus a powerful
tool for the study of their intrinsic properties. Interestingly,
in view of the conventional Elliot mechanism \cite{Zutic:04} for
spin relaxation in metals, one expects a relatively long spin
relaxation length (several $\mu m$), because of the expected low
spin-orbit coupling. This makes carbon nanotubes potentially
attractive for device applications.

The main problem encountered in previous studies of electrical
spin injection in NTs was to find ferromagnetic metals which can
contact reliably the NTs, with a low ohmic device resistance
\cite{Tsukagoshi:99,Kim:00,Nygard:00}. A low device resistance is
not a priori needed in a macroscopic spin valve, where the
conductance is controlled by the relative orientation of the
magnetization of two ferromagnetic electrodes around an insulating
barrier. However, transparent ferromagnetic contacts on NTs are
essential for the study of spin dependent transport at low
temperatures, to avoid quenching transport because of charging
effects.

In this letter, we present a scheme for contacting MWNTs with a
ferromagnetic alloy, Pd$_{1-x}$Ni$_{x}$, with x $\approx 0.7$.
Shape anisotropy is used for controlling the coercive field of the
ferromagnetic contacts. This scheme allows to achieve contact
resistances of, on average, {$30$ $k\Omega$} at room temperature.
The minimum devices resistance measured so far is {$5.6$
$k\Omega$}. The yield of devices with resistances below {$100$
$k\Omega$}, at {$300$ $K$}, is around $50\%$. In the linear
conductance regime, we find that the resistance switches
hysteretically when sweeping the magnetic field. The relative
difference between the resistance in the antiparallel (AP)
orientation and the parallel (P) orientation is about $2\%$.

Giant paramagnetism is a well-known feature of Pd and few magnetic
impurities added to its matrix can drive it into the ferromagnetic
state \cite{Beille:75}. Therefore, as Pd alone makes
quasi-adiabatic contacts on NTs \cite{Dai:03}, ferromagnetic Pd
alloys are expected to keep the same contacting properties as Pd,
provided the concentration of magnetic impurities is low enough.
However, as the spin signal is proportional to $P^{2}$, $P$ being
the spin polarization of the alloy \cite{Juillere:75}, the
concentration should not be too small to ensure that the current
which is driven in the MWNTs is enough spin polarized. As we will
see below, the contacting properties of Pd$_{1-x}$Ni$_{x}$ on NTs
remain very similar to pure Pd even in the case of high Ni
concentration. Therefore, we chose to use the alloy in the
concentrated limit (x$\approx$0.7) to ensure high enough spin
polarization of the source-drain current.

 The MWNTs used in this work are
grown by arc discharge and stored as a suspension in chloroform.
We first pre-pattern Au bonding pads and alignment marks on a
thermally oxidized highly doped Si (resistivity of {$\approx$ $5$
$m \Omega.cm$} at {$300$ $K$}),
used as back-gate (SiO$_{2}$ thickness $%
\approx 400nm$). We spread the MWNTs on this substrate and
localize them under a SEM (Philips FEG XL30 or LEO). Using
conventional e-beam lithography techniques, we write and develop
the structure shown after
lift-off on figure \ref{fig:photo}. In a vacuum system  with a base pressure of about $%
5.10^{-8}$ $mbar$, we first deposit a layer of Pd$_{1-x}$Ni$_{x}$ with $x$ $%
\approx$ $0.7$ of $600$ \AA , at a pressure of about $10^{-7}$
$mbar$. We use angle evaporation to obtain isolated ferromagnetic
electrodes. We finally deposit $400$ {\AA}  of Pd to connect the
device to the pre-patterned Au bonding pads (not shown on figure
\ref{fig:photo}). This Pd/PdNi bilayer is also evaporated on a
bare substrate placed nearby in order to characterize the alloy by
SQUID magnetometry and RBS (Rutherford Backscattering
Spectrometry). For all the samples for which we could study spin
injection (6 samples), the spacing of the ferromagnetic pads was
either $1$ $\mu m$ or $500$ $nm$. As shown on figure
\ref{fig:photo}, the ferromagnetic electrodes have different
shapes. This is to achieve different coercive fields for the two
electrodes, by shape anisotropy, in order to produce a spin valve.
Typical dimensions are {$14$ $\mu m$ $ \times$ $0.1$ $\mu m$} and
{$3$ $\mu m$$\times$ $0.5$ $\mu m$} for the left and the right
electrode respectively.

On figure \ref{fig:histo}a and b, we show the magnetic
characterization of a thin film of $600$ \AA\ of
Pd$_{1-x}$Ni$_{x}$ with x $\approx $ $0.7$ under $400$ \AA\ of Pd.
The magnetic field dependence of the magnetization displays a
hysteresis loop with a coercive field of {$50$ $mT$} (the field is
perpendicular to the layer). The magnetization saturates at around
$0.25$ $\mu _{B}$ per total number of atoms and decreases rapidly
above $270$ $K$.\ Note that, although this is enough to study spin
transport below $100$ $K$, the Curie temperature and the
saturation magnetization are $50\%$ smaller than the known bulk
characteristics for this Ni concentration \cite{Beille:75}. We
think that this might be due to partial oxidation of the Ni during
evaporation. Evaporating the alloy at a lower pressure could allow
to achieve room temperature ferromagnetic contacts. The Ni
concentration in the Pd matrix is measured on the same thin film
by RBS.

Figure \ref{fig:histo}c shows the histogram of the device
resistances. On the hundred of NTs contacted so far, we could
contact successfully $46$ of them with a device resistance below
$100$ $k\Omega $. As shown on the histogram, the{\it \ average}
device resistance of these $46$ samples is around
$30$ $k\Omega$ at $300$ $K$. The lowest device resistance was found to be $5.6$ $k\Omega $ at $%
300$ $K$ and the most probable one is $20$ $k\Omega $. All these
resistances are measured for a gate voltage $V_{G}=0.0$ $V$ in the
linear regime. At $2$ $K$, the linear resistance remains typically
below $100$ $k\Omega $ when sweeping the gate, which allows to
study spin transport.

The dependence of the linear resistance $dV/dI$ of a device versus
an applied magnetic field $H$ for two different gate voltages
$V_{G}=2.00$ $V$ and $V_{G}=0.00$ $V$ is shown on figure
\ref{fig:magneto}. In order to take advantage of shape anisotropy,
the field is kept parallel to the axis of the ferromagnetic pads.
The overall behavior is a decrease of the resistance as one
increases the magnetic field, as previously reported
\cite{Bachtold:99,Schoen:99}. In addition, for $V_{G}=2.00$ $V$,
the resistance displays a hysteretic behavior. Around $0$ $mT$, it
gradually increases further upon reversing the sign of the
magnetic field and switches to a lower value around $100$ $mT$. As
expected for a spin valve, the two curves $dV/dI(H)$ obtained when
sweeping down or up match at high field and are roughly
mirror-symmetric. Therefore, when reversing the sign of the
magnetic field, the region between $0$ $mT$ and $60$ $mT$
corresponds to an antiparallel (AP) orientation of the
magnetizations of the electrodes, whereas all the other regions of
field correspond to the parallel (P) orientation. We define the
$TMR$ as

\[
TMR=2\frac{R_{AP}-R_{P}}{R_{P}+R_{AP}}
\]

where $R_{AP}$ is the resistance in the AP orientation at $50$
$mT$ and $R_{P}$ is the resistance in the P orientation at the
same field. For $V_{G}=2.0$ $V$, the $TMR$ is positive, around
2.05\%. Even though the exact spin polarization of the alloy is
not known, one can roughly estimate it comparing the magnetization
of the actual alloy with that of pure Ni. Taking the known value
for the spin polarization $P_{Ni}$ of Ni\cite{Tedrow:94}, one
obtains $P_{PdNi}\approx \mu_{PdNi}P_{Ni}/\mu_{Ni}\approx
0.25*23/0.6=9.58\%$. This spin polarization would yield a TMR of
$1.85$ $\%$ for a tunnel junction within the simple Julli\`ere's
model \cite{Juillere:75}. Although this amplitude is in reasonable
agreement with our measurements, this comparison probably
underestimates spin-dependent and/or energy dependent scattering
in the nanotube. For example, charging effects could be important.
They are indeed observed in the spin independent part of the $R$
vs $V_{G}$ characteristic.

The resistance change of about $2$ $\%$ measured for $V_{G}=2.0$
$V$ could also be accounted for by a change of about $50$ $mT$ of
local stray magnetic field arising from the ferromagnetic pads
contacting the nanotube. For ruling out this spurious effect, one
can define the sensitivity to external local magnetic fields of
the nanotube-device as the slope of the $R$ vs $H$ curve when
there is no magnetic switching. For $V_{G}=2.0$ $V$ , it is
$0.037$ $\%/mT$ ($14$ $\Omega/mT$) and for $V_{G}=0.0$ $V$ , it is
$0.018$ $\%/mT$ ($6$ $\Omega/mT$). Thus, a change in the stray
fields of $50mT$ would indeed change the resistance at $V_{G}=2.0$
$V$ of about $2$ $\%$ but would also change the resistance at
$V_{G}=0.0$ $V$ by $1$ $\%$. However, as shown on figure
\ref{fig:magneto}, there is a hysteresis lower than $0.1$ $\%$ in
the $R$ vs $H$ curve for $V_{G}=0.0$ $V$, whereas a hysteresis of
of $2$ $\%$ is present for $V_{G}=2.0$ $V$. We can therefore rule
out stray magnetic field effects from the ferromagnetic pads, as
they should be independent of the gate voltage. The electronic
current flowing through the tube is spin-polarized. Note also that
figure \ref{fig:magneto} shows that the TMR is gate controlled.
This gate dependence is presently not understood and will be
studied in subsequent papers.

In conclusion, we have demonstrated reliable contacting and spin
injection in MWNTs with transparent contacts. Using a Pd$_{1-x}$Ni$_{x}$ alloy with x$%
\approx $ 0.7, we can have device resistances as low as $5.6$ $
k\Omega$ at $300$ $K$. At $2$ $K$, we observe a TMR of about
$2\%$. We think that this contacting scheme will allow extensive
studies of spin effects in NTs in the 0D or the 1D regime and can
be used in principle for device applications.

$Acknowledgments$. We acknowledge fruitful discussions with R.
Allenspach, W. Belzig and A. Cottet. We thank F. Lalu and M.
Aprili for the RBS measurements. We thank Lazlo Forr\`o for the
MWNTs. This work is supported by the DIENOW RTN network, the NCCR
on nanoscience and by the Swiss National Fundation.

\pagebreak

\pagebreak

FIG. \ref{fig:photo} A SEM picture of a device. The
Pd$_{0.3}$Ni$_{0.7}$ electrodes have different shapes, $ 14$ $\mu
m\times $ $0.1$ $\mu m$ and $3$ $\mu m\times$ $ 0.5$ $\mu m$ for
the left and the right electrode respectively. They are spaced by
$1$ $\mu$m. The black arrows indicate the direction of the
magnetization in the AP or the P orientation

FIG.\ref{fig:histo} a. Temperature dependence of the magnetization
of a thin film of Pd$_{0.3}$Ni$_{0.7}$ of 600 {\AA} coated by
500{\AA} of Pd, obtained while evaporating the contacts on the NT.
The Curie temperature of the alloy is around $270$ $K$. b.
Magnetic field dependence of the magnetization of the
Pd$_{0.3}$Ni$_{0.7}$ film at $T=2.7$ $K$. As expected, a
hysteresis is observed. The saturation magnetization is about
$0.25$ $\mu_{B}$.c. Histogram of the contacting properties of PdNi
on MWNTs at 300K. The mean resistance is $30$ $k\Omega$ and the
most probable one is $20$ $k\Omega$

FIG. \ref{fig:magneto} Magnetic field dependence of the linear
resistance at $1.85$ $K$ as a function of an in-plane magnetic
field parallel to the axis of the ferromagnetic electrodes, for
gate voltages $V_{G}=2.0$ $V$ and $V_{G}=0.0$ $V$. A hysteretic
behavior characteristic of a spin valve is observed for
$V_{G}=2.0$ $V$


\begin{references}
\bibitem{Datta:90}S. Datta and B. Das, Appl. Phys. Lett. {\bf 56}, 665 (1990).
\bibitem{Egger:01} L. Balents and R. Egger, Phys. Rev. B {\bf 64},
 035310 (2001).
\bibitem{Dekker:98}S.J. Tans, R.M. Verschueren and C. Dekker, Nature,{\bf 394}, 761 (1998).
\bibitem{Smalley:99}M. Bockrath, D.H. Cobden, J. Lu, A.G. Rinzler, R.E. Smalley, L. Balents, P.L. McEuen, Nature,{\bf 397}, 598 (1999).
\bibitem{Zutic:04}I. Zutic, J. Fabian and S. Das Sarma, Rev. Mod.
Phys., {\bf 76}, 323 (2004).
\bibitem{Tsukagoshi:99}K. Tsukagoshi, B.W. Alphenaar and H. Ago, Nature,{\bf 401}, 572 (1999).
\bibitem{Kim:00}J.R. Kim, H.M. So, J.J. Kim and J. Kim, Phys.
 Rev. B {\bf 66}, 233401 (2002).
\bibitem{Nygard:00}B. Zhao, I. M\"{o}nch, H. Vinzelberg, T. M\"{u}hl, and C. M. Schneider, Appl. Phys. Lett. {\bf 80}, 3144 (2002).
\bibitem{Dai:03} A. Javey et al., Nature {\bf 424}, 654 (2003). We have developed independently the contacting
 with Pd and PdNi. See B. Babic, T. Kontos and C. Sch\"{o}nenberger, Cond-mat 0406626 and B.
 Babic, Proc. of the XVII Winterschool, Kirchberg, Austria,
 (2003).
\bibitem{Juillere:75} M. Julli\`ere, Phys. Lett. {\bf54A}, 225
(1975).
\bibitem{Beille:75}J. Beille, Ph. D. thesis, Universit\'e Joseph
Fourier, Grenoble, 1975.
\bibitem{Bachtold:99} A. Bachtold, Christoph Strunk, Jean-Paul Salvetat, Jean-Marc Bonard, Laszló Forró, Thomas Nussbaumer and Christian Schönenberger
, Nature,{\bf 397}, 673 (1999).
\bibitem{Schoen:99} C. Sch\"{o}nenberger, A. Bachtold, C. Strunk, and J.-P. Salvetat
, Appl. Phys. A, {\bf 69}, 283, (1999).
\bibitem{Tedrow:94}R. Meservey and P.M. Tedrow, Phys.
 Rep. B {\bf 238}, 173 (1994).
\end{references}
\end{document}